# Financing Artificial Intelligence Infrastructure: Mapping AI Infrastructure Investment and Compute Governance Across Africa

**Kai-Hsin Hung[1,2], Sumaya Nur Adan[3], Krupa Suchak[4], Armita Sadeghian Barzoki[5], Kofi Yeboah[6], Mohammad Amir Anwar[7]**

[1]HEC Montreal
[2]International Development Research Centre
[3]University of Oxford
[4] Dataspire
[5] Toronto Metropolitan University
[6]Mozilla Foundation
[7]University of Edinburgh and University of Johannesburgh

kai-hsin.hung@hec.ca, sumaya.adan@cs.ox.ac.uk, info@dataspires.com,
armita.sadeghian@torontomu.ca, kofi@mozillafoundation.org, mohammad.anwar@ed.ac.uk

**Abstract**

Artificial intelligence depends on large-scale compute resources and their supporting infrastructure. However, AI governance debates treat compute primarily as a technical input rather than as an outcome of investment, ownership, and financial control. This paper examines AI infrastructure investment flows across Africa through a systematic analysis of 46 publicly announced projects totalling USD $12.7 billion between 2019 and 2025. Using a value chain framework, we analyze who invests in AI-relevant infrastructure and where investments concentrate. Our findings reveal a highly concentrated landscape dominated by global data center operators, hyperscale technology firms, and development finance institutions, clustering in South Africa, Kenya, Nigeria, and Egypt. We introduce *asymmetrical interdependence* to describe a structural condition in which capital and physical infrastructure account for 73% of total funding while control remains concentrated in the compute layer among a small number of global technology firms. We argue that compute governance must account for capital flows, ownership, and control, not only geographic access, because these dynamics shape AI compute equity. Infrastructure presence is necessary but insufficient for meaningful governance capacity.

## 1 INTRODUCTION

Artificial intelligence (AI) is shaped not only by technological innovation but also by patterns of capital investment and infrastructure finance. Data centers, cloud computing platforms, and high-performance computing (HPC) facilities have become the material foundations of AI capability, and who builds, owns, and controls these facilities has direct consequences for who can develop, deploy, and govern AI systems (Day and Bang, 2026; Panday and Gruenwald, 2025). As AI systems become increasingly dependent on large-scale computational capacity, control over compute resources has emerged as a critical dimension of AI governance and political and economic contestation globally.

Africa sits at the sharpest edge of this problem. The continent holds less than one percent of global AI compute resources, making it one of the starkest examples of the emerging AI compute divide (Besiroglu et al., 2024; Davies and Vipra, 2024). Only 5% of AI talent on the continent has access to compute resources (Tsado and Lee, 2024; ADCA, 2024; AI4D Africa, 2024), and the region is routinely characterized as a compute desert defined by absence and limited hyperscale infrastructure (Lehdonvirta et al., 2024). Our research disrupts that narrative. Between 2019 and 2025, at least USD $12.7 billion was committed across 46 publicly announced AI infrastructure projects on the conti-

nent, with investment accelerating sharply after 2022. Africa is not a compute desert. It is becoming a site of significant and contested AI infrastructure investment.

The critical question is therefore not whether investment is arriving, but what governance capacity it produces and for whom. Infrastructure presence and governance capacity are not the same thing. A hyperscale data center financed by foreign private equity, operated by a global platform, and dependent on proprietary GPU architecture can expand local connectivity while concentrating control over AI capability and dependency in a small number of external actors. Infrastructure expansion can embed local digital ecosystems within externally controlled technological and financial networks (Velasco et al., 2026), and understanding these dynamics requires examining not only where AI infrastructure is built, but who financed it and who ultimately controls it.

This paper develops that account through a systematic analysis of 46 AI infrastructure investment projects totalling USD $12.7 billion across Africa between 2019 and 2025. Using a value chain framework, we address two research questions. First, who invests in AI infrastructure in Africa? Second, where do these investments concentrate, and what governance implications follow?

Our findings reveal what we term *asymmetrical interdependence*: a structural condition in which investment, ownership, and control are distributed unevenly across the AI infrastructure value chain in ways that systematically disadvantage local governance capacity. Capital and physical infrastructure account for 73% of total investment, and while African state actors, pan-African development finance institutions (DFI), and regionally headquartered actors participate in these layers, ownership remains substantially concentrated among foreign private equity firms and global infrastructure operators. Control is most concentrated in the compute layer, which despite representing only 15.6% of total investment, global technology firms dominate it entirely. Microsoft, Huawei, and NVIDIA determine how infrastructure is used by controlling cloud platforms, compute access, pricing models, and technical standards. Even when infrastructure is financed and built locally, the ability to use AI systems is governed through external platforms, centralizing control among a small number of actors and creating structural dependencies for downstream users. Those who invest are not always those who own, and those who own do not necessarily control how AI systems are used or by whom.

Therefore, we use the term *compute equity* to describe the conditions under which actors, regions, and communities can not only access AI compute infrastructure but also participate meaningfully in its ownership, governance, and the distribution of value it generates. This extends the compute divide literature, which focuses primarily on geographic access gaps, to incorporate questions of control, accountability, and local capability that physical infrastructure presence alone cannot resolve.

This matters for the AIES and compute governance communities in three concrete ways. First, physical infrastructure presence is not a reliable proxy for regulatory capacity: hosting infrastructure does not equal governing it. Second, compute equity cannot be assessed by mapping data center locations alone, because the compute layer where cloud platforms, pricing, and technical standards are controlled is entirely absent of African state actors or regionally headquartered platforms, meaning the terms on which infrastructure becomes usable AI capability are set wholly externally. Third, geographic clustering compounds rather than corrects existing inequalities, concentrating investment where connectivity, scientific and knowledge advantages already exist and reproducing the inequalities infrastructure investment is often framed as solving.

The paper proceeds as follows. Section 2 reviews the global AI investment landscape, compute governance literature, and Africa's position within longer histories of digital foreign direct investment (FDI) and inequality. Section 3 describes our dataset and methodology. Sections 4 and 5 present findings on who invests and where. Section 6 discusses governance implications, limitations and future research.

## 2 BACKGROUND & RELATED WORKS

### 2.1 The Global AI Infrastructure Investment Supercycle

AI increasingly depends on large-scale physical infrastructure, and the financial flows underpinning that infrastructure are growing at an unprecedented pace. The AI sector is currently experiencing what analysts describe as an investment supercycle, with estimates suggesting up to $3–5 trillion in capital expenditure required globally by 2030 (Bank of England, 2025; JLL, 2026). Global data center capacity is projected to grow by roughly 14 percent annually, doubling within five years (JLL, 2026). The four largest U.S. hyperscalers, Microsoft, Meta, Alphabet, and Amazon, are expected to invest approximately $650 billion in capital expenditures in 2026 alone, up from $430 billion in 2025 (Day and Bang, 2026). AI-related infrastructure accounted for roughly $270 billion in announced FDI flows in 2025, with the majority concentrated in advanced economies including the United States, France, and South Korea (UNCTAD, 2026). The geography of AI infrastructure investment is therefore highly uneven from the outset, reproducing and extending existing patterns of concentration.

### 2.2 Compute as a Governance Problem

Recent scholarship has moved beyond mapping infrastructure distribution toward understanding compute as a governance problem. Compute resources are expensive, geographically concentrated, and controlled by a small number of firms, creating structural asymmetries in the ability of actors, organizations, and regions to participate meaningfully in AI development (Lehdonvirta, 2022; Lehdonvirta et al., 2024). This has led researchers to emphasize compute governance as a distinct analytical lens, highlighting how access to and control over AI infrastructure shapes the development, deployment, and regulation of AI systems (Sastry et al., 2024). Unlike many other components of AI systems, data centers are measurable physical infrastructures with specific geographic locations, making them potential focal points for policy intervention, industrial strategy, and regulatory oversight (Heim and Koessler, 2024). Within emerging national AI stack frameworks, compute hardware and data centers are increasingly treated as strategic assets underpinning digital economic growth, technological sovereignty, and the broader diffusion of AI applications (Bianchini et al., 2022; OECD, 2023; Hooker, 2020).

### 2.3 Africa as Compute Desert & Compute Equity

Africa sits at the sharpest edge of these global asymmetries. The continent currently holds less than one percent of global compute capacity, and only five percent of AI talent on the continent has access to compute resources (Tsado and Lee, 2024; ADCA, 2024). This has led to characterizations of Africa as a 'compute desert,' defined by the absence of hyperscale infrastructure and limited access to public AI computational resources (Lehdonvirta et al., 2024). Scholars have drawn on broader theoretical frameworks to situate this within longer histories of technological and economic marginalization. Manuel Castells described Africa as a potential black hole of informational capitalism, systematically excluded from the networks through which value and knowledge circulate globally (Castells, 2000). More recent work has raised concerns that new waves of AI infrastructure investment risk reproducing rather than reversing these patterns, particularly when projects are financed, owned, and operated primarily by foreign actors (Velasco et al., 2026).

Existing research has largely mapped the geographic distribution of hyperscale data centers and compute capacity (Davies and Vipra, 2024; Lehdonvirta et al., 2024; Sevilla et al., 2022). These studies provide important insights into global disparities, but they tend to treat compute capacity as a static resource. In practice, the expansion of AI infrastructure is driven by ongoing investment decisions, financing arrangements, and partnerships between technology firms, infrastructure investors, telecommunications providers, and various finance institutions (Gelvanovska-Garcia et al., 2024; Velasco et al., 2026). This paper addresses that gap by examining not just where AI infrastructure exists in Africa, but who finances it, who owns it, and who ultimately controls it.

This distinction motivates our use of the term *compute equity*. Where the compute divide literature focuses primarily on geographic access gaps, compute equity describes the conditions under which actors, regions, and communities can not only access AI compute infrastructure but also participate meaningfully in its ownership, governance, and the distribution of value it generates (Lehdonvirta et al., 2024; Sastry et al., 2024). Compute equity is not achieved when infrastructure is built in a region. It is achieved when the actors in that region can determine how that infrastructure is governed, how it is priced, and for whose benefit it operates. This framing is what this paper applies to Africa's emerging AI infrastructure and investment landscape.

### 2.4 FDI and Digital Infrastructure in Africa: Historical Patterns and Persistent Inequalities

Africa's engagement with foreign direct investment (FDI) in digital infrastructure follows a long-established pattern in which capital flows into the continent without necessarily generating local ownership, capability, or distributional benefit. Africa remains the least connected continent, with only 38% of its population having internet access as of 2024, a disparity rooted not simply in technological lag but in the structural conditions under which digital and energy infrastructure has been built and governed (UNCTAD, 2025). Foreign corporations dominate the continent's digital backbone. Undersea cables owned by firms such as Google and Meta carry almost 90% of Africa's internet traffic, functioning as gatekeepers of continental connectivity and mirroring colonial trade routes in their design, encircling Africa rather than integrating it (Global Voices, 2025). The resulting hub-and-spoke model passes costs downward, with the average cost of 1GB of data in Africa reaching 5.7% of monthly income, nearly three times the affordability threshold.

The aggregate investment figures obscure this structural reality. In 2024, FDI flows to Africa rose sharply to $97 billion, up from $55 billion in 2023. However, this headline figure rests substantially on a single Egyptian mega-project and does not reflect broad-based digital investment across the continent (UNCTAD, 2025). The relationship between FDI, digital infrastructure, and inclusion is conditional rather than straightforward. Research on Sub-Saharan Africa finds that the relationship between digital infrastructure, and inclusive growth is conditional and context-dependent rather than straightforward, with digital infrastructure showing statistically insignificant effects on inclusive growth as a moderator of aid flows (Ofori et al., 2024). High-speed internet connectivity does attract further FDI, particularly in finance, technology, and health services, but connectivity alone does not resolve structural exclusion when ownership and governance remain externally concentrated (Mensah and Traore, 2022).

### 2.5 The AI Divide: Compute Access, Digital Inclusion, and the Risk of Reproducing Inequality

The arrival of AI infrastructure investment in Africa does not automatically alter these structural conditions and in some respects, risks deepening them. Over 83% of AI startup funding in 2025 went to Kenya, Nigeria, South Africa, and Egypt, leaving much of the continent marginal to the flows shaping AI development (Mwangi, 2025). Africa's AI market is projected to reach $16.53 billion by 2030, yet the continent holds less than 1% of global compute capacity, and only 5% of AI talent on the continent has meaningful access to compute resources (Tsado and Lee, 2024; ADCA, 2024).

The digital divide risks becoming sharper rather than narrower in the age of AI. The urban-rural gap remains stark, with nearly 60% of urban residents in countries like Kenya and Nigeria online, while rural access often falls below 20%, and over 600 million Africans still lack reliable electricity (Disrupt Africa, 2025). Gender exclusion compounds this further, with women severely underrepresented in ICT and AI fields across the continent. These are not peripheral concerns but sectoral and structural constraints on whether AI infrastructure investment generates genuine local capability or simply expands the reach of externally governed actors.

This is precisely the tension that compute governance frameworks, and to a lesser extend digital sovereignty, must address. Physical infrastructure build-out can enable digital economy growth, foster inclusion, and catalyze investment at unprecedented scale. But as the historical record of digital FDI in Africa demonstrates, infrastructure presence does not guarantee equitable outcomes when ownership, pricing, and governance remain concentrated upstream. The current wave of AI infrastructure investment puts Castells' characterization directly to the test. Whether it marks a genuine structural shift or a reproduction of earlier dependency patterns depends not on the volume of capital flowing in, but on who controls the infrastructure once it is built.

## 3 METHODOLOGY

The study draws on a dataset of publicly announced investments in AI-relevant digital infrastructure across Africa between 2019 and 2025. Data collection was conducted through a systematic review process using Covidence for screening and coding, and Google Sheets for structured data extraction and tracking. The dataset captures private and public investment announcements related to data centres, cloud computing infrastructure, and associated facilities enabling AI workloads.

We conducted a comprehensive search using Factiva alongside specialized industry and regional sources, including Data Center Dynamics and AllAfrica. The search yielded an initial pool of 1,225 records down to 289 records with about one-third from media, a third from industry reports, and a third from peer-reviewed articles. Following four iterative rounds of individual and team-based screening and adjudication, 46 sources were retained for inclusion. The temporal scope was anchored to begin in 2019 to capture pre-pandemic baseline investment activity and extends through the end of 2025.

Search and inclusion criteria were structured around two keyword clusters: (1) investment-related terms (financing, investment, foreign direct investment, capital investment) and (2) infrastructure-related terms (data cent*, supercomput*, AI infrastructure, high-performance comput*). This approach ensured systematic identification of projects directly linked to the expansion of AI-relevant compute capacity.

Only projects directly related to AI-relevant compute infrastructure, primarily data centres, cloud regions, and HPC facilities, were included. Each investment was coded using a structured framework capturing four dimensions: actors (who finances and receives), geography (where it is located), infrastructure type (what is built), and investor and firm identity, with estimation procedures applied for multi-country projects.

Given the lack of standardized metrics and mandatory disclosure of compute capacity, their related investment flows were used as a proxy for in-coming infrastructure expansion. This approach reflects broader opacity in the sector, including inconsistent reporting by hyperscalers and infrastructure operators (Sevilla et al. 2023; Lehdonvirta et al. 2025). Our final dataset includes 46 projects, refined through iterative coding and team-based validation. Where financial data were incomplete or aggregated, values were disaggregated and conservatively estimated using company disclosures, industry reports, and comparable benchmarks, with adjustments for partial reporting, phased investments, and currency variation. Estimates are treated as indicative and supported by contextual insights from industry sources and practitioner input.

# 4 WHO: FOUR KEY AI INFRASTRUCTURE VALUE CHAIN INVESTORS & ROLES

Across 46 AI infrastructure investment projects between 2019 and 2025, totaling USD $12.675 billion, AI infrastructure can be understood as layered value chains made up of different types of investors with interconnected roles. These roles can be grouped into four core layers: (1) the capital and financing layer, which funds and shapes where infrastructure is developed; (2) the physical infrastructure layer, which builds and operates data centers and facilities; (3) the compute infrastructure layer, which provides cloud platforms, software, and AI computing capabilities; and lastly (4) the connectivity and network layer, which links users to these systems through telecommunications networks.

Capital providers such as private equity firms, sovereign investors, development finance institutions (DFI), and commercial banks determine what projects are funded, where investments are directed, and how infrastructure is owned. These financial flows enable the construction of physical infrastructure, where data center operators and regional platforms build and operate the facilities that host AI systems. On top of this foundation, global technology firms and AI compute providers turn infrastructure into usable AI services by running cloud platforms, deploying GPU-based computing systems, and providing the software tools needed to build and use AI. At the same time, regional and local telecommunications operators expand fiber and mobile networks to connect users and businesses to these systems, making it possible to access AI services across different regions.

These four layered are important because it reveals that investment, ownership, and control are distributed across the

AI infrastructure value chains, see Table 1 below. While most of the capital is concentrated in financing and physical infrastructure, control over how AI systems are accessed and used is bottlenecked at the compute layer, where a small number of global firms set the terms of participation. Connectivity providers play a critical role in extending access but operate downstream with more limited influence over value creation and capture. Understanding AI infrastructure through these investor roles highlights how different actors shape the development of AI systems and helps explain why power and value are not evenly distributed across the value chain.

Table 1: **Investor categories across four in AI infrastructure value chains ordered by total investment from 2019-2025**

| Digital Infrastructure Layers | Investor Categories | Representative Actors | Core Function in AI Infrastructure Value Chain | # Projects | Total Investment (USD $M) | % of Total |
|---|---|---|---|---|---|---|
| **1. Capital & Financing Layer** | Private Equity, Sovereign, Development Financial Institutions, Commercial Banks | Berkshire Partners LLC; Actis LLP; Helios Investment Partners LLP; Group 42 Holding Ltd. (G42), United Arab Emirates; Abu Dhabi Exports Office (ADEX), part of the Abu Dhabi Fund for Development; Ports, Customs and Free Zone Corporation (PCFC), Government of Dubai; International Finance Corporation (IFC); World Bank Group; Africa50 Infrastructure Investment Platform; British International Investment (BII); United States International Development Finance Corporation (U.S. DFC); Export-Import Bank of the United States (EXIM); Rand Merchant Bank (a division of FirstRand Limited); Standard Bank Group Limited; Nedbank Group Limited | Finance infrastructure through equity, debt, concessional funding, and sovereign investment, shaping where and how infrastructure is built | 17 | 4,514 | 35.61% |
| **2. Physical Infrastructure Layer** | Global Data Center Operators & Regional Infrastructure Platforms | Digital Realty Trust Inc.; Equinix Inc.; Vantage Data Centers LLC; Teraco Data Environments; Cassava Technologies (Africa Data Centres); Raxio Group; Wingu Africa; Texaf; ST Digital; N+ONE Datacenters; Oceinde; MainOne; Olkaria Eco-Cloud | Build, own, and operate data centers and physical facilities that host compute, storage, and interconnection systems | 13 | 4,729 | 37.31% |
| **3. Compute Infrastructure Layer** | Global Technology Firms & Hyperscale Cloud Providers | Microsoft Corporation; Huawei Technologies Co., Ltd.; NVIDIA Corporation | Provide cloud platforms, AI compute (including GPUs), and software ecosystems that run on top of physical infrastructure | 6 | 1,977 | 15.60% |
| **4. Connectivity & Network Layer** | Sub-regional & Local Telecommunications Operators | Airtel Africa; MTN Group Limited; Safaricom; Vodacom Group Limited (including Vodacom Mozambique); Orange S.A. (including Orange Botswana); Mauritius Telecom Ltd.; Botswana Fibre Networks Proprietary Limited (BoFiNet); Federal Government of Somalia (Ministry of Communications and Technology) | Provide fiber, mobile networks, and edge infrastructure connecting users, data centers, and cloud systems | 10 | 1,455 | 11.48% |
| | | | **Total** | **46** | **12,675** | **100%** |

*Note: Investment figures are in USD millions. Percentages may not sum to 100% due to rounding. Multi-country projects are disaggregated and conservatively estimated. Figures are indicative.*

## 4.1 Capital & Financing Layer

The capital and financing layer is where AI infrastructure begins. It brings together a mix of investors operating at different scales, including global private equity firms such as Berkshire Partners, Actis, and Helios; sovereign and state-linked actors like G42, ADEX, PCFC, and EXIM; development finance institutions including IFC, Africa50, BII, and U.S. DFC; and regional and local commercial banks such as Rand Merchant Bank, Standard Bank, and Nedbank. These actors collectively determine which projects move forward, where infrastructure is built, and who ultimately owns it.

Across the dataset, this layer accounts for $4.514 billion, or 35.61% of total investment, across 17 projects. Major transactions include Berkshire Partners' $1 billion acquisition of Teraco, which scaled one of Africa's leading data center platforms, and the $1 billion Microsoft–G42 partnership in Kenya, highlighting the growing role of sovereign-linked capital. Additional investments such as Actis' $250 million platform in Nigeria, IFC's $100 million investment in Raxio, and EXIM's $100 million financing in Côte d'Ivoire demonstrate how global capital is combined with development finance and local banking systems to support infrastructure expansion.

The Berkshire Partners' acquisition of a majority stake in Teraco Data Environment in early 2019 provided the initial capital required to expand the company's interconnection campuses in South Africa (Berkshire Partners 2019). Similarly, infrastructure investors have supported the development of regional data center platforms designed to scale across multiple markets. By providing strategic capital and financial resources, infrastructure investors play a crucial role in enabling infrastructure operators to scale data center capacity across emerging markets.

Commercial banks and infrastructure lenders form another important component of AI infrastructure financing. Institutions such as Rand Merchant Bank, Standard Bank, and Nedbank collectively account for approximately $169 million across two projects (1.6% of total investment). These institutions typically provide project loans, refinancing packages, and syndicated credit facilities that support the construction and expansion of data center infrastructure.

Development finance institutions (DFIs) provide another important source of capital for digital infrastructure projects, particularly in markets where private investment alone may be insufficient. Institutions such as the International Finance Corporation (IFC), Africa50 Infrastructure Investment Platform, British International Investment (BII, formerly CDC Group), and the U.S. Development Finance Corporation account for approximately $345 million across six projects (1.3.% of total investment).

What is particularly important is how this layer is structured across scales. Global investors bring large pools of capital and shape platform-level ownership, regional and multilateral institutions reduce risk and enable market entry, and local banks provide on-the-ground financing and liquidity. Together, they do more than fund infrastructure, they shape who controls it, how it is governed, and how risks and returns are distributed, increasingly influenced by both market incentives and geopolitical priorities.

### 4.2. Physical Infrastructure Layer

The physical infrastructure layer provides the backbone of AI systems, including data centers, power systems, cooling infrastructure, and interconnection hubs. This layer is led by global operators such as Digital Realty, Equinix, Vantage Data Centers, and Teraco, alongside regional platforms like Cassava Technologies (Africa Data Centres), Raxio, and Wingu Africa, and smaller local or specialized operators such as Texaf, ST Digital, and N+ONE Datacenters. Together, these actors build and operate the facilities that host compute, storage, and network systems.

The physical infrastructure layer represents the largest share of investment, with $4.729 billion, or 37.31% of total investment, across 15 projects. Key developments include the $1.725 billion acquisition of Teraco by Digital Realty, the $1 billion Vantage hyperscale campus in Johannesburg, and Teraco's $419 million expansion. Additional projects such as Equinix's $320 million acquisition of MainOne in Nigeria and Kenya's $626 million Project Eagle data center show how infrastructure is expanding both through global platform integration and regional growth.

For example, Airtel Africa's Nxtra data center projects in Nigeria and Kenya aim to expand regional cloud capacity while integrating new facilities into the company's telecommunications network (Sehloho 2025). Similarly, Safaricom's data center developments in Kenya support the local hosting of digital services and enterprise cloud applications. By strengthening regional connectivity and hosting capacity, telecommunications operators help extend access to cloud and AI services across emerging markets.

The structure of this layer reflects a clear division of roles across scales. Global operators bring capital, technical expertise, and integration into international networks, regional platforms expand infrastructure across multiple African markets, and local operators fill gaps in emerging or underserved areas. This layered structure drives the formation of infrastructure hubs, particularly in South Africa, Kenya, and Nigeria, while also reinforcing uneven development. Ultimately, this layer determines where AI infrastructure exists but remains dependent on compute providers to drive demand and utilization.

### 4.3 Compute Infrastructure Layer

The compute layer is where AI infrastructure becomes usable. It includes cloud platforms, GPU systems, and software environments that allow organizations to train models, run applications, and deploy AI systems. The key actors here are global technology firms such as Microsoft and Huawei, alongside AI compute providers like NVIDIA, and firms such as Visa that deploy specialized data infrastructure. Unlike other layers, this one is overwhelmingly dominated by global actors, with limited regional or local presence.

Investment in this layer totals $1.977 billion, or 15.60% of the total, across 6 projects. Major examples include Microsoft's Azure regions in South Africa (~$300M) and its $1 billion partnership with G42 in Kenya, as well as Huawei's cloud deployments in South Africa and Egypt (~$300M each). The $720 million Cassava–NVIDIA AI Factory highlights the expansion of GPU-based AI infrastructure across multiple markets and signals a shift toward more specialized compute capacity.

Despite its smaller share of investment, this layer holds the most concentrated control. These actors determine how compute is accessed, priced, and used through platforms, APIs, and technical standards. NVIDIA's role is especially important as it provides the underlying hardware and software architecture without owning infrastructure directly. This creates a system where AI infrastructure can be locally financed and built but remains globally controlled at the level of compute and capability, reinforcing a concentration of power among a small number of firms.

### 4.4 Connectivity & Network Layer

The connectivity layer ensures that AI infrastructure can be accessed and used. It includes fiber networks, mobile infrastructure, and edge systems that connect users to data centers and cloud platforms. This layer is led primarily by regional and local telecommunications operators such as Airtel Africa, MTN, Safaricom, Vodacom, Orange, and Mauritius Telecom, alongside state-linked or national infrastructure providers such as Botswana's BoFiNet and Somalia's Ministry of Communications and Technology. Compared to other layers, global actors play a more limited direct role here.

This layer accounts for $1.455 billion, or 11.48% of total investment, across 10 projects. Major investments include Airtel's Nxtra data centers in Nigeria (~$340M) and Kenya (~$150M), MTN's $235 million facility in Nigeria, and Mauritius Telecom's $434 million regional connectivity hub project. Smaller-scale developments, such as Safaricom's Limuru data center in Kenya and Vodacom Mozambique's facility, extend connectivity into local markets and support broader infrastructure coverage.

What distinguishes this layer is its strong regional and local orientation. Regional telecom operators manage cross-border infrastructure and large-scale networks, while local and state-linked actors ensure last-mile connectivity and national integration. Within the AI infrastructure value chain, regional infrastructure platforms operate within the regional hosting layer, providing local access points for cloud services and digital platforms. By developing infrastructure in smaller markets and secondary cities, these platforms help extend digital infrastructure beyond major regional hubs and support the growth of local digital ecosystems.

This makes the connectivity layer essential for expanding access and enabling participation in digital economies. However, it remains structurally downstream, as it does not control compute platforms or core infrastructure. As a result, it plays a critical role in diffusion and access, but captures less value and influence compared to upstream layers.

## 5. WHERE: GEOGRAPHIC DISTRIBUTION OF AI INFRASTUCTURE INVESTMENT INTO FOUR REGIONAL HUBS

The geography of AI infrastructure investment across Africa reveals a highly concentrated hub-and-gateway structure, in which a small number of metropolitan regions host the majority of hyperscale data-centre capacity while a wider network of secondary nodes extends connectivity and cloud access across the continent. Analysis of the investment covering 2019–2025 shows that a limited group of regional hubs accounts for most of the capital deployment. In particular, South Africa (~50%), Kenya (~20%), Nigeria (~10%), Egypt (~2%), and multi-regional (10-12%) host the largest clusters of infrastructure investment and together account for most of the hyperscale and large colocation facilities.

These core hubs are complemented by a second tier of regional infrastructure nodes, including Ghana, Côte d'Ivoire, Mozambique, and Ethiopia, where smaller facilities support localized cloud hosting and interconnection services. A third tier of connectivity gateways, such as Djibouti, Mauritius, Cape Verde, Togo, and Réunion, plays a strategic role in linking continental networks to global submarine cable systems and telecommunications backbones.

### 5.1 South Africa: Primary Hyperscale Hub for the Continent

South Africa represents the largest and most mature AI infrastructure ecosystem on the African continent, serving as the principal hyperscale hub for Southern Africa and the continent more broadly. Approximately half of all capital investment is concentrated in South Africa, reflecting its central role in hosting large-scale cloud infrastructure and colocation facilities.

The metropolitan areas of Johannesburg, Cape Town, and Germiston contain the largest concentration of data centre campuses and interconnection facilities in Africa. Major projects include the expansion of Teraco's data-centre campuses, the construction of the Equinix JN1 facility, and the Vantage JNB2 hyperscale campus. In addition, the deployment of Microsoft Azure cloud regions in Johannesburg and Cape Town established the first hyperscale public cloud presence on the continent, enabling regional access to enterprise cloud services and AI computing platforms.

Figure 1: **Regional AI Infrastructure Hubs, Secondary Nodes, and Regional Gateways and Connectivity Points**

These investments reflect several structural advantages that position South Africa as a continental infrastructure hub. The country benefits from relatively advanced telecommunications networks, established colocation ecosystems, and multiple submarine cable landing stations connecting Africa to global internet backbones. Major infrastructure operators such as Digital Realty and Equinix have entered the market through acquisitions and new construction projects, reinforcing Johannesburg's role as the primary interconnection point for cloud providers, telecommunications carriers, and content delivery networks serving Southern Africa.

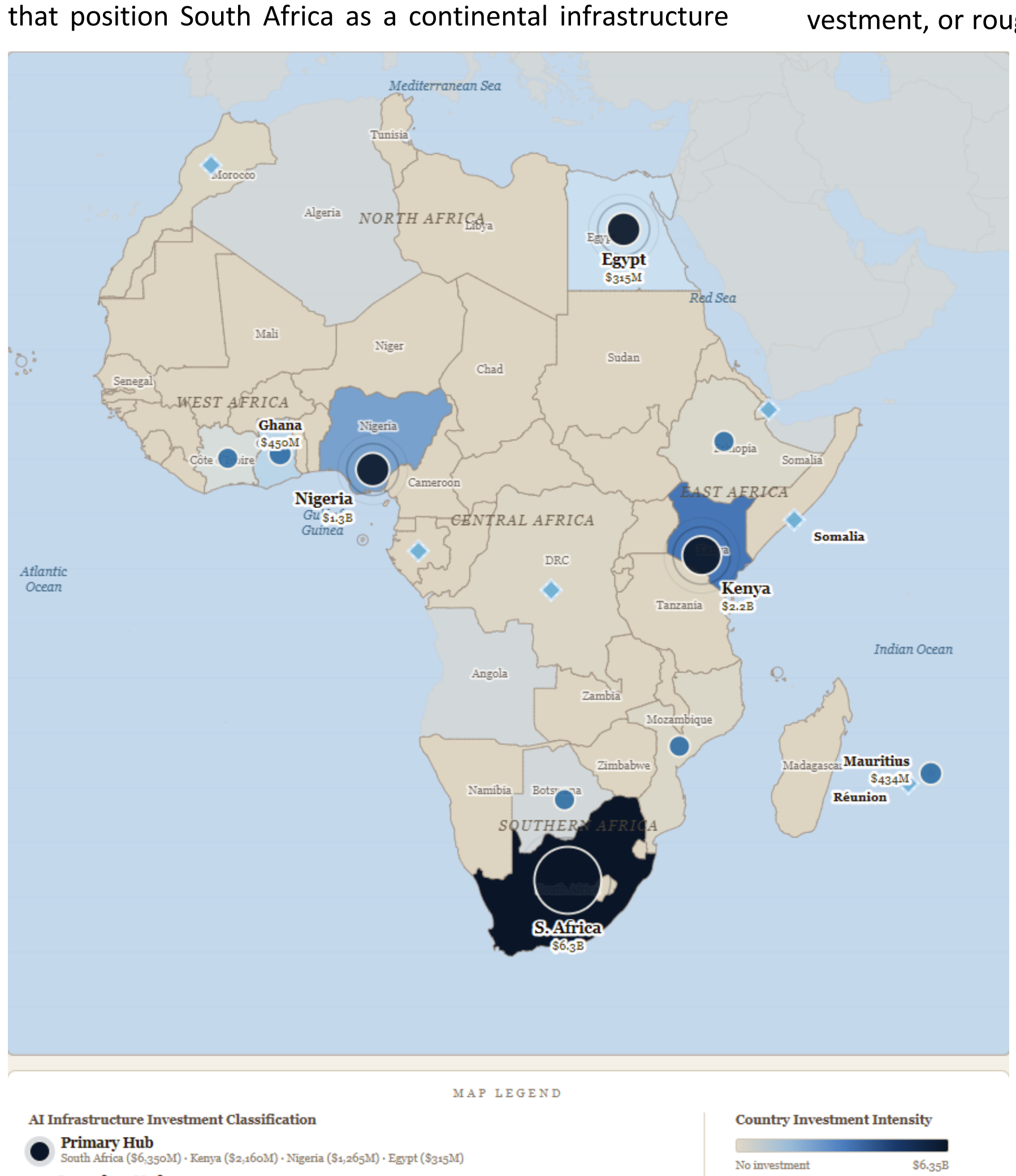

## 5.2 Kenya: East African AI Hub

Kenya has emerged as the leading AI infrastructure hub in East Africa, accounting for approximately $2.16 billion in investment, or roughly 19% of total capital deployment. The Nairobi metropolitan region and nearby locations such as Naivasha and Kiambu host several major infrastructure projects including the Project Eagle hyperscale facility, the Microsoft-G42 AI infrastructure initiative, and the Safaricom Limuru data centre.

Kenya's importance within the regional digital economy is closely linked to its position within East Africa's telecommunications and connectivity networks. Nairobi hosts key internet exchange points and connects to multiple submarine fiber-optic cable systems along the Indian Ocean coast. These networks enable Kenya to function as a regional distribution hub for cloud services and digital platforms.

As hyperscale cloud providers expand infrastructure across emerging markets, Kenya increasingly serves as a gateway through which AI-enabled services and cloud computing resources are distributed to neighboring markets, including Uganda, Tanzania, and Rwanda. The expansion of hyperscale data centre infrastructure and connectivity networks therefore positions Kenya as the core node of the East African digital infrastructure ecosystem.

## 5.3 Nigeria: West African Hub

Nigeria represents the largest digital economy in West Africa and has become a critical infrastructure gateway for cloud and data-centre services across the region. Infrastructure investment in Nigeria totals approximately $1.15 billion, representing roughly 10% of total investment. Most infrastructure projects are concentrated in Lagos; the country's financial and telecommunications centre. Major developments include the Nxtra Lagos hyperscale campus developed by Airtel Africa, the MTN Nigeria data-centre project, and the Equinix acquisition of MainOne, which integrated West African infrastructure into the global Equinix colocation network.

Nigeria's role as a regional infrastructure hub is supported by its large population, expanding digital services sector, and extensive telecommunications networks. Lagos hosts multiple submarine cable landing stations, including high-capacity international fiber systems connecting West Africa to Europe and the Americas. By locating data-centre infrastructure near these connectivity nodes, operators enable low-latency access to cloud platforms and enterprise applications across the region. As a result, Nigeria functions as the primary cloud gateway for West Africa, enabling enterprises and digital platforms to host services locally while maintaining connections to global cloud ecosystems.

### 5.4 Egypt: North African Hub

Egypt has emerged as an important digital infrastructure gateway linking Africa, Europe, and the Middle East, with approximately $315 million in AI and cloud infrastructure investment in the dataset, representing roughly 3% of total capital deployment. Morocco is an emerging hub alongside Egypt in Northern Africa. A central driver of Egypt's infrastructure development has been the expansion of Huawei's regional cloud and AI infrastructure. In 2022, Huawei launched the Huawei Cairo Cloud Region, supported by an investment of approximately $300 million, marking one of the company's largest cloud infrastructure deployments in Africa. This facility forms part of Huawei's broader strategy to expand cloud computing services, AI platforms, and enterprise digital infrastructure across African markets. The Cairo cloud region provides services including AI development tools, enterprise cloud platforms, and data hosting capacity for regional governments and businesses.

Additional investment includes the Raya Data Center project developed in partnership with Africa50, representing approximately $15 million in infrastructure investment. Together, these projects contribute to the expansion of Egypt's data center and cloud infrastructure ecosystem. Egypt's strategic importance is closely tied to its geographic position along major international fiber-optic cable routes connecting Europe and Asia through the Mediterranean and Red Seas. A large share of global internet traffic between Europe and Asia passes through Egyptian territory via submarine cable systems and terrestrial fiber networks. As a result, data centers located in Cairo serve not only domestic demand but also regional markets across North Africa and the Middle East.

### 5.5 Secondary Infrastructure Nodes and Connectivity Gateways

Beyond the four primary regional hubs, a set of secondary infrastructure nodes supports the broader distribution of cloud and AI infrastructure across the continent. These markets extend regional cloud access and data hosting capacity and can be grouped geographically across Southern, West, North, and East Africa, as well as key cross-regional connectivity gateways.

Multi-regional and pan-African investments represent a distinct layer of capital deployment that does not anchor in a single geography but instead operates across multiple national markets simultaneously. These investments estimated at roughly $1.2B–$1.5 billion are typically led by a mix of regional infrastructure platforms, DFIs, and strategic technology partnerships.

In Southern Africa, Mozambique (~$54M) and Botswana (~$20M) host smaller data-centre projects supporting regional connectivity and enterprise cloud hosting across Southern Africa. In East Africa, Ethiopia ($50M) is developing early-stage digital infrastructure through the Wingu Africa ICT Park data centre in Addis Ababa, reflecting growing demand for domestic data hosting and enterprise cloud services.

In West Africa, Côte d'Ivoire (~$129M) and Ghana ($50M) serve as emerging regional infrastructure nodes. Côte d'Ivoire's infrastructure developments around Grand-Bassam, including the MainOne data centre and EXIM-backed national data centre initiative, support digital infrastructure expansion across Francophone West Africa. Ghana's PAIX data centre in Accra provides carrier-neutral colocation infrastructure connecting telecommunications networks and cloud providers across the region.

In Central Africa, the Democratic Republic of Congo (DRC) (~$20M) hosts the Open Access Data Centres (OADC) facility in Kinshasa, developed with investment from Texaf. This project represents one of the first carrier-neutral data-centre facilities in the country and aims to improve domestic data hosting capacity while strengthening connectivity across Central African telecommunications networks. Gabon (~$9M) hosts smaller infrastructure investments supporting enterprise hosting capacity and regional telecommunications interconnection.

In North Africa, Morocco ($15M) hosts a smaller colocation facility near Nouaceur, contributing to the expansion of regional data hosting capacity and strengthening North African connectivity with European infrastructure networks.

Finally, a number of Indian Ocean countries and locations function primarily as connectivity gateways linking regional networks to global digital infrastructure systems. Mauritius ($434M) represents the largest secondary infrastructure investment. Mauritius Telecom's hyperscale data centre and submarine cable expansion aim to position the island as a regional interconnection hub linking Africa and Asia. Djibouti (~$75M across regional projects) has emerged as a major submarine cable landing hub connect-ing Africa to Europe and Asia, while Réunion (~$10M) supports regional telecommunications and data hosting capacity in the western Indian Ocean. These secondary infrastructure nodes and connectivity gateways extend the reach of cloud and AI infrastructure beyond the primary hyperscale hubs, enabling regional markets to integrate into broader global digital and AI infrastructure networks.

# 6 DISCUSSIONS: WHY THIS MATTERS FOR AIES & COMPUTE GOVERNANCE?

The findings here challenge two oversimplifications shaping debates about AI infrastructure in Africa. The first is that Africa remains a compute desert. At $12.7 billion across 46 projects, concentrated largely after 2022, this is a substantive investment footprint that did not exist six years ago. The second is that investment straightforwardly translates into capability, equity, or governance participation. The evidence suggests the relationship is far more conditional, and the gap between infrastructure presence and governance capacity deserves serious attention.

Four patterns define this landscape: high capital concentration in a small number of projects, strong geographic clustering in a few established digital hubs, accelerating investment activity after 2022, and a stakeholder mix of hyperscale technology firms, infrastructure operators, and development finance institutions with distinct and not always aligned interests.

Together these patterns produce what we term *asymmetrical interdependence*. Capital and physical infrastructure, where African-adjacent actors have some presence, account for 73% of total funding. Yet control sits in the compute layer, which at 15.6% of total investment is dominated entirely by global technology firms. Microsoft, Huawei, and NVIDIA collectively determine how AI systems are accessed, priced, and governed across the continent. Infrastructure can be locally financed and built while remaining externally governed. Those who invest are not always those who own, and ownership does not necessarily confer control over how AI systems are used.

This has direct implications for compute governance. Existing frameworks focus on whether compute presence enables states to regulate AI within their jurisdictions (Lehdonvirta et al., 2024; Sastry et al., 2024). Our findings suggest location is necessary but insufficient. Hosting a hyperscale data center financed by foreign private equity and dependent on proprietary GPU architecture does not automatically confer regulatory leverage. South Africa, Kenya, Nigeria, and Egypt are gaining infrastructure presence, but accountability remains concentrated upstream among global operators outside these jurisdictions. Geographic clustering compounds this, reinforcing rather than redistributing existing inequalities, while grid reliability, pricing opacity, and limited local stakeholder agency remain persistent structural constraints.

The promise of compute is real. Infrastructure build-out enables digital economy growth, fosters inclusion, and has catalyzed investment at unprecedented scale. But value capture remains skewed toward foreign hyperscalers, investors, and governments, and execution risk, grid instability, and limited local agency are ongoing threats. Realizing this investment moment requires not just more infrastructure, but more equitable ownership and greater local control over how compute is governed.

## 6.1 AI Investments, Ownership, and Control

Investment, ownership, and control in Africa's AI infrastructure reveal a clear structural asymmetry. Investment is concentrated in capital and physical layers, accounting for roughly 73% of total funding, reflecting the capital-intensive nature of data center construction and large-scale platform financing. Ownership in these layers is held predominantly by global entities such as Digital Realty, Equinix, and international investment funds, creating a separation between local asset location and external governance. Control, despite representing only 15.6% of total investment, is concentrated in the compute layer, where firms like Microsoft, Huawei, and NVIDIA govern cloud platforms, pricing models, and technical standards. Even where infrastructure is financed and built locally, the ability to deploy AI systems is mediated through external platforms. The result is that control over AI capability is centralized among a small number of global actors, creating structural dependencies that reinforce asymmetries across the entire value chain.

## 6.2 Limitations & Future Research

Africa's AI infrastructure investment landscape is expanding rapidly, but growth alone does not guarantee equitable or locally governed compute capacity. This paper has mapped $12.7 billion in investment across 46 projects, revealing a structurally asymmetrical value chain in which capital flows into physical infrastructure while control remains concentrated among a small number of global technology firms. For compute governance frameworks, the central implication is clear: physical presence is not a reliable proxy for regulatory capacity or local capability.

Several limitations shape the scope of these findings. The dataset relies on publicly announced investments, meaning undisclosed or private transactions are not captured. Investment value is used as a proxy for infrastructure capacity given the opacity of compute reporting across the sector. Notably, the energy infrastructure layer, which is critical to data center operations and AI workloads across the continent, falls outside the scope of this analysis and represents an important gap.

Future research should examine energy dependencies and power grid infrastructure as a distinct governance layer, track ownership changes over time as projects mature, and investigate how local stakeholder agency evolves as investment scales. Understanding who ultimately benefits from Africa's compute expansion remains an open, conditional and open question.

## Acknowledgments

This study was supported from a research award from the UK's FCDO-IDRC AI for Development (AI4D) program hosted at Canada's International Development Research Centre (IDRC). The views expressed herein do not necessarily represent those of IDRC or its Board of Governors.

The authors gratefully acknowledge the helpful comments from Gillian Dowie and Anindya Chaudhuri and from the three anonymous reviewers. We also presented at Deep Learning Indaba 2025 in Kigali, Rwanda.